\documentstyle[preprint,aps]{revtex}
\begin{document}
\draft

\title{PROBING ANOMALOUS TRIPLE BOSON \\
       VERTICES AT FUTURE $e^{+}e^{-}$ COLLIDERS}

\author{Pat Kalyniak and Paul Madsen}
\address{Ottawa-Carleton Institute for Physics, Physics Department, Carleton
University \\
1125 Colonel By Drive, Ottawa, Canada K1S 5B6}

\author{Nita Sinha and Rahul Sinha}
\address{Institute of Mathematical Sciences, CIT Campus Taramani,\\
Madras, India, 600113}
\maketitle

\begin{abstract}
{We explore the detection potential of the four lepton production
processes $e^{+}e^{-} \rightarrow l^{+} \nu l^{\prime -}\overline{\nu}$ for
anomalous contributions to the triple boson vertices at proposed future
high energy colliders with center-of-mass energies of 500 GeV and 1 TeV.
The predicted bounds are of the order of a few percent for
the $CP$-even couplings $\kappa_{V}$ (V=$\gamma$,Z) at the higher energy; we
show
that these  limits can be improved by as much as a factor
of two through suitable phase space cuts. A polarized beam facility,
with its ability to access helicity information, could provide
constraints on the vertices significantly tighter than those achievable
from an analysis of total cross-section alone.
The bounds on the $CP$-odd coupling
$\tilde{\lambda}_{V}$ approach the indirect bounds from neutron electric
dipole measurements while those on $\tilde{\kappa}_{V}$ are much looser.
The asymmetries in experimental observables produced by
such an explicitly $CP$ violating triple vertex contribution are
seen to be below the expected level of statistical precision of approximately
$1.5 \%$; asymmetries in the individual contributing helicity amplitudes
might however be detectable.}
\end{abstract}
\pacs{PACS number(s): 13.10.+q, 14.80.Er}
\newpage
\section{Introduction}

It will be possible to directly measure the Triple Boson Vertices (TBV)
at future high energy $e^{+} e^{-}$ colliders like the CERN Large Electron
Positron Collider II (LEPII) \cite{davier,kane}
and the Next Linear Collider
(NLC) \cite{hawaii,desy,finland}. These gauge boson
self-couplings are a key prediction of the non-abelian
$SU(2)_{L} \times U(1)_{Y}$ electroweak theory and, as yet, are only loosely
constrained by indirect
loop contributions and measurements of the
 $p \overline{p} \rightarrow W^{\pm} \gamma, W^{\pm}Z$,
and $W^{+}W^{-}$ processes. Both the
D0 and CDF collaborations have now looked for $W\gamma$ \cite{abe,ellison},
$WZ$ \cite{fuess}, and $W^{+}W^{-}$ \cite{fuess,abachi} production in the data
from the 1A run (1992-93). Present 95 \% CL experimental limits are
$-2.3 <\Delta \kappa_{\gamma} < 2.2$ (CDF)
and $-1.6 < \Delta \kappa_{\gamma} < 1.8$ (D0)
from $p \overline{p} \rightarrow W\gamma$ and
$-1.0 < \Delta \kappa_{V} < 1.1$ (CDF)
and $-2.6 < \Delta \kappa_{V} <2.8 $ (D0) from
$p \overline{p} \rightarrow WW, WZ$. Two analyses of recent CLEO data
\cite{cleo} on the process b$\rightarrow s \gamma$ determine consistent
limits of $-1.44 < \Delta \kappa_{\gamma} < 1.5$ \cite{mackhe} and
$-0.41 < \Delta \kappa_{\gamma} < 1.22$ \cite{martinez}.
While these experimental bounds are compatible with the Standard Model (SM),
they
are still too weak to be considered a precision test of the theory.

The couplings of the W to the neutral gauge bosons $\gamma$ and Z can
be described in general by an effective Lagrangian with seven parameters.
A standard parametrization of the vertices is \cite{hag}
\begin{eqnarray}
\label{lag}
L_{VWW}/g_{V} & = & ig_{1}^{V}(W^{\dag}_{\mu \nu}W^{\mu}V^{\nu}
- W^{\dag}_{\mu}V_{\nu}W^{\mu \nu}) + i\kappa_{V}W^{\dag}_{\mu}
W_{\nu}V^{\mu \nu}  \nonumber \\
		& + & \frac{i \lambda _{V}}{M^{2}_{W}}W^{\dag}
_{\lambda \mu}W^{\mu}_{\nu}V^{\nu \lambda} - f_{4}^{V}W^{\dag}
_{\mu}W_{\nu}(\partial ^{\mu}V^{\nu} + \partial ^{\nu}V^{\mu})
		    \nonumber \\
		& + & f_{5}^{V}\epsilon ^{\mu \nu \rho \sigma}
(W^{\dag}_{\mu} \stackrel{\leftrightarrow}{\partial}_{\rho}W_{\nu})V_{\sigma}
+ i\tilde{\kappa}_{V}W^{\dag}_{\mu}W_{\nu}\tilde{V}^{\mu \nu}
		    \nonumber \\
		& + & \frac{i\tilde{\lambda}_{V}}{m^{2}_{W}}
W^{\dag}_{\lambda \mu}W^{\mu}_{\nu}\tilde{V}^{\nu \lambda}
		    \nonumber
\end{eqnarray}
where V represents either the photon or Z field and the overall
couplings are $g_{\gamma} = e$ and $g_{Z} = e \cot \theta _{W}$.

Of the seven coupling parameters, $g_{1}^{V}, \kappa _{V}, \lambda _{V}$,
and $f_{5}^{V}$ parametrize $CP$ respecting effective Lagrangian terms;
their tree-level standard model values are
\begin{displaymath}
g_{1}^{V} = ~~\kappa _{V} = 1, ~~\lambda _{V} = f_{5}^{V} = 0
\end{displaymath}

Most previous work on anomalous TBV contributions has concentrated on these
$CP$-even couplings \cite{hag,godfrey,kaly}. For a recent review see
\cite{dpf}.
Predicted detection bounds on these couplings at the Large Hadron Collider
(LHC)
and an NLC with $e^{+}e^{-}$ center of mass energy of 500 GeV or greater
are at the percent level and better when a variety of processes is considered.
This degree of sensitivity should be
sufficient to measure the TBV at the level of Standard Model loop corrections
\cite{couture} and certain extensions to the SM \cite{susy}.

The couplings $f_{4}^{V}, \tilde{\kappa}_{V}$, and $\tilde{\lambda}_{V}$
parametrize vertex contributions that violate $CP$ invariance.
Since $CP$ violation was first discovered in the neutral
$K^o - \overline{K}^o$ system, a satisfactory explanation of its origin
has been lacking. Within the
$SU(2)_{L} \times U(1)_{Y}$ framework of the
SM, $CP$ violation occurs via the Kobayashi-Maskawa phase in the CKM matrix.
Since the
CKM matrix lies in the quark sector, a $CP$ violating
contribution to a leptonic process
will appear to first order at the two-loop level in the SM; these
$CP$-odd couplings are therefore zero at tree and one-loop level. One-loop
effects are
however possible in extensions to the SM; they manifest themselves in
non-zero $f_{4}^{V}, \tilde{\kappa}_{V}$, and $\tilde{\lambda}_{V}$
of the order of 10$^{-2}$ or smaller due to the loop suppression
\cite{einhorn,susy}.
Such an explicitly $CP$ violating
contribution to the WW$\gamma$ vertex is strongly constrained by neutron
electric
dipole measurements to be less than $\sim (10^{-4})$ \cite{neut};
the SU(2)
symmetry then implies that a WWZ contribution should be similar in magnitude.
Nevertheless, because these experimental constraints on a
$CP$ violating contribution to the TBV
are indirect, they are no substitute for a direct search.

We have previously considered the purely leptonic processes
\begin{equation}
\label{proc}
e^{+}e^{-} \rightarrow l^{+} \nu l^{- \prime } \overline{\nu}
\end{equation}
with all possible charged lepton combinations in the final state as candidates
for
measuring the triple boson vertex. The details of that
examination are given in \cite{kaly}. Detection limits
for the $CP$-even couplings $\kappa_{V}$, achievable through measurement
of the total cross-section at energies of 500 GeV and 1 TeV were there
determined.
At $\sqrt{s}$ = 500 GeV, we found
\begin{eqnarray*}
-2.5\%(\mu \tau) <\Delta \kappa_{\gamma}<+8.0\% (ee)
\nonumber \\
-4.5\%(\mu e) <\Delta \kappa _{Z} <+8.0\% (\mu e) \nonumber
\end{eqnarray*}
and at $\sqrt{s}$ = 1 TeV,
\begin{eqnarray*}
-1.0 \% (\mu \tau) <\Delta \kappa _{\gamma} <+3.5 \% (\mu \tau,\mu e)
\nonumber \\
-1.5 \% (\mu \tau) <  \Delta \kappa _{Z} <     +2.5 \% (\mu \tau)
\nonumber
\end{eqnarray*}
where the parentheses indicate which of the charged lepton combinations
provide these tightest bounds.

In this paper, we extend our consideration of the processes of
Eq.(\ref{proc}). The paper consists of two parts. In the first,
we discuss means by which the above limits on the $CP$-even couplings
$\kappa_{V}$ might be improved.
We discuss in Section \ref{pol} the
advantages offered by accessing the helicity information through a
polarized beam facility and
present in Section \ref{even} the improved limits achievable through the
restriction of certain angular variables' phase space.
In the second part of the paper, we examine the potential for
detection of non-standard values for the $CP$-odd couplings
$\tilde{\kappa}_{V}$ and $\tilde{\lambda}_{V}$.
We present in Section \ref{detlim} our results for detection
limits on the couplings $\tilde{\kappa}_{V}$ and $\tilde{\lambda}_{V}$,
derived from both total and differential cross-section analyses.
In Section \ref{asym} we consider
the possibility of asymmetries in certain $CP$ odd variables as providing
more sensitive indicators of $CP$ violation than does the cross-section.
We summarize in Section \ref{con}.

\section{Improving Limits for $CP$-even couplings $\kappa_{V}$}
\label{first}

\subsection{Polarized Beams}
\label{pol}

In the energy ranges of LEP and higher, in contrast with
the left-right symmetric physics of the QED dominated lower energies,
helicity effects are expected to be important. These polarization effects
might prove useful
in the measurement of the triple boson vertices WW$\gamma$ and WWZ.
The possibility of using polarized $e^{+}e^{-}$ beams to access the helicity
information in a measurement of the TBV has been examined
previously \cite{davier,perkins,bilchak,likhoded}. At LEPII energies,
polarized beams  are not expected to provide a substantial gain in
sensitivity, but might help to disentangle different anomalous contributions
\cite{davier,perkins}. At higher energies, significantly
improved limits on the couplings could be achieved with initial beam
polarization \cite{bilchak,likhoded}.

Individual helicity amplitudes for a given process can differ
in their dependence, both in form and magnitude, on the
triple boson vertices.
The contributions of different
helicity amplitudes to the total cross-section can therefore provide
constraints on $\kappa _{\gamma}$ and $\kappa_{Z}$ that are complementary
to those obtained from the total cross-section.
For instance, the
$\mu^{+} \tau^{-} $ channel has contributions from only the
$(+-+-)$ and $(-++-)$
helicity amplitudes. Our convention is to list the helicities of the
four charged fermions as $(e^{+},e^{-},l^{+},l^{'-})$.
The $(+-+-)$ amplitude involves the neutrino exchange diagram of
W-pair production, the $(-++-)$ amplitude
does not.
This diagram dominates the $(+-+-)$ amplitude as well as the total
cross-section
but unfortunately does not contain the trilinear couplings. The neutrino
exchange diagram can therefore be thought of as a $\kappa _{V}$-independent
``background'' to
the contributions of the triple boson vertex containing diagrams.
The $(-++-)$ amplitude, and resulting cross-section, is free
of the large $\kappa _{V}$-independent contribution of this diagram, and
has therefore a relatively greater sensitivity to deviations
$\Delta \kappa _{V}$. Without the dominant t-channel neutrino
contribution however, the cross-section
$\sigma_{-++-}$ is much smaller than that of $\sigma_{+-+-}$
($\sigma_{-++-} \sim 10^{-2}\sigma_{+-+-}$),
with consequently poorer statistical sensitivity.
It therefore requires analysis to determine
whether the improved sensitivity to $\Delta \kappa _{V}$ of
$\sigma_{-++-}$ will be sufficient to counter the loss of statistics,
and ultimately provide more stringent limits.

To illustrate the scale of the possible improvements in the detection limits
for the
deviations $\Delta \kappa _{V}$, we examine the individual helicity amplitude
contributions to the total cross-section for the $\mu^{+} \tau^{-}$
channel at $\sqrt{s}$ = 500 GeV. The $\mu^{+} \tau^{-}$ channel is chosen
because,
with only two helicity
amplitudes contributing, it provides the simplest
demonstration of the principle. The other final state configurations present
complications due to the larger number of contributing helicity amplitudes.
Without the
ability to measure the polarization of the final state fermions, what
would actually be measured are the four different possible combinations of
initial polarizations, $++, +-, -+,$ and $--$, some of which would be sums over
separate helicity amplitudes.

As before, we explore the sensitivity of the cross-sections to variations in
$\kappa _{\gamma}$ and $\kappa _{Z}$, but now also
 determine the percentage
contribution of each helicity state to the total cross-section. We
convert these percentages into effective component cross-sections, from which
helicity detection limits on $\kappa _{\gamma}$ and $\kappa _{Z}$ can be
determined.
For reference, the Standard Model total cross-section for the
$\mu^{+} \tau^{-}$ process
at $\sqrt{s}=500$ GeV, including also
the charge conjugate channel $\mu^{-} \tau^{+}$, is $\sigma_{SM}$=0.0684 pb.
The individual helicity state contributions to this total are determined to
be, $\sigma _{SM}^{(+-+-)}$=0.0676 pb and $\sigma _{SM}^{(-++-)}$=0.0008 pb.

We show the results of the analysis in
Fig. 1. The concentric solid
lines are the $\pm 2 \sigma$ contours for the $(+-+-)$ helicity state;
the ``disk'' they define is
very similar to that of the total cross-section. The four parallel diagonal
dotted lines are the $\pm 2 \sigma$ contours corresponding to the
$(-++-)$ amplitude. This distinctive form is a consequence of
the cancellations
that occur between the photon and Z contributions to the $(-++-)$ amplitude
for equal pairings of $\kappa _{\gamma}$ and $\kappa _{Z}$.
The SM Feynman rules for the couplings of $\gamma$ and Z to massive leptons,
plus the different coupling strengths of the two triple boson vertices
$\gamma$WW and ZWW, ensure that the $\kappa$-dependence of the $(-++-)$
amplitude can be schematically written as
\begin{equation}
\label{5.3}
M_{(-++-)}= \left(\frac{\kappa _{\gamma}-1}{s}-\frac{\kappa _{Z}-1}
{s-M_{Z}^{2}}\right)(A) + (B)
\end{equation}

Thus, for
equal values of non-standard $\kappa _{\gamma}$ and $\kappa _{Z}$ and
$s \gg M_{Z}^{2}$,
the different contributions of the photon and Z bosons will nearly cancel.

Neither the $(+-+-)$
disk region or the $(-++-)$ bands independently offer significantly tighter
constraints on
$\kappa _{\gamma}$ and $\kappa _{Z}$ than did the unpolarized cross-section;
the intersection of the former with the latter does however severely restrict
the allowed domain. What had
been a relatively large set of $\kappa _{\gamma},\kappa _{Z}$ pairings
statistically indistinguishable from the Standard Model prediction is
reduced to a significantly smaller union of two separate regions.
Polarization measurements could therefore, in the specific case of the
$\mu^{+} \tau^{-}$ final state, provide significantly tighter constraints
on $\kappa _{\gamma}$ and $\kappa _{Z}$ than would the unpolarized
cross-section. For instance, if we vary the couplings individually, keeping the
other
at its SM value, we determine limits
\begin{eqnarray*}
0.99 < \kappa _{Z} < 1.01 & & \\
0.99 < \kappa_{\gamma} < 1.01 & ~~~\mbox{and} ~~~& 1.08 < \kappa_{\gamma} <
1.09
\end{eqnarray*}
which are a significant improvement on the bounds achievable from the
unpolarized analysis.

For large center-of-mass energies, the cancellation between the photon and Z
terms when $\kappa_{\gamma}=\kappa _{Z}$
results in a $(-++-)$ helicity
amplitude contribution of less than about $1\%$ of the total cross-section.
When $\kappa _{\gamma} \neq \kappa _{Z}$ however, this amplitude can
contribute a much larger portion of the total, the above cancellation being
destroyed.
A small percentage contribution of $(-++-)$ to the total cross-section
is therefore
characteristic of equal values for $\kappa _{\gamma}$ and $\kappa _{Z}$ and
thus provides a characteristic ``signature'' of such pairings.
More than merely helping to distinguish between the cases of equal or unequal
$\kappa _{\gamma}$ and $\kappa _{Z}$, the relative contribution of the
$(-++-)$ amplitude could potentially serve as a characteristic ``fingerprint''
of a particular region of the $\kappa _{\gamma},\kappa _{Z}$ grid.
 A measurement of a
non-standard cross-section can, in general, be attributed to any of an infinite
set of
pairings of $\kappa _{\gamma}$ and $\kappa _{Z}$, these points lying on a
contour of constant cross-section. Although indistinguishable by a simple
measurement of total cross-section, these pairings will, in general,
have very
different characteristic helicity contributions. Thus, once a cross-section
measurement had indicated a non-standard $\kappa _{\gamma},\kappa _{Z}$
pairing, the relative contributions of the helicity amplitudes to the
total cross-section could differentiate between the
sets of possible pairings responsible.

Similar analyses were performed for the other final state lepton pairings.
The bounds from the other helicity amplitudes that contribute to these
more complicated processes were generally seen to be too loose to
provide any further constraint on the couplings.

\subsection{Sensitivity Enhancing Cuts}
\label{even}

In this Section, we examine the results of making various phase space cuts.
We show in Fig. 2
the distributions for the angular variable $\cos \theta _{\tau ^-}$
(where $\theta _{\tau ^-}$ is the angle between the outgoing
lepton and the incoming positron)
for the $e^+ e^- \rightarrow \mu ^+ \nu \tau ^- \overline{\nu}$
process at $\sqrt{s}=500$ GeV. The solid line corresponds to the
Standard Model values, the dashed line to $\kappa _{\gamma}=\kappa _{Z}
=0.9$, and the dotted line to $\kappa _{\gamma}=\kappa _{Z}=1.1$.

The strong peak in the distribution at  $\cos \theta_{\tau^{-}} \simeq -1$
is a consequence of the t-channel
neutrino exchange diagram of W pair production. This behaviour is repeated
for the three other processes with different final state lepton
configurations, but is less marked. There, the many extra
diagrams beyond W pair production dilute the effect of the t-channel
neutrino diagram.

For non-standard pairings of $\kappa _{\gamma}$ and $\kappa _{Z}$, these
angular distributions are generally somewhat enhanced in the regions of
phase space away from this peak. Since the t-channel
neutrino exchange diagram does not contain the triple boson vertex, the peak
is relatively insensitive to the vertex couplings.
This localization of the $\kappa$ sensitivity to the ``non-peak''
regions of the
$\theta_{l^{-}}$ phase space suggests the potential for
maximizing the $\kappa$ sensitivity by cutting on these variables to
exclude the large non-sensitive peak contributions, and
isolate the $\kappa$ dependent plateau regions. Because, by making
such a cut, we lose a significant portion of our total cross-section,
we must distinguish between improving the {\it physical} sensitivity,
as a percentage deviation from the SM total cross-section, and the
{\it experimental} sensitivity, defined as the potential for detection
of anomalous couplings.

The angular variables are not unique in their localization of sensitivity
to $\kappa _{V}$. Another potential observable is $M_{l^{+} l^{\prime-}}$,
the invariant mass of the
outgoing charged lepton-antilepton pair. The sensitivity
to non-standard $\kappa_{V}$ is located predominantly in the middle
region of the
$M_{\mu \tau}$ distribution for the $\mu^{+} \tau^{-}$ process
at $\sqrt{s}=500$ GeV. A cut
such as 30 GeV $< M_{\mu \tau} < 430$ GeV
would exclude the $\kappa$-insensitive contributions from the extreme low and
high invariant mass regions of phase space.
We can therefore consider
making combined cuts, both an angular and invariant mass cut, in hopes of
improving sensitivity.

To simplify matters, we restrict our study to the case
$\kappa_{\gamma}
=\kappa_{Z}$. Any limits derived with the angular cuts in place,
although more constraining than those
obtained by making no assumptions about the relationship between
$\kappa_{\gamma}$ and $\kappa_{Z}$, will still
demonstrate the scale of the improvements in sensitivity
possible.

In Table I, for the $\mu \tau$ channel at
$\sqrt{s} = 500 $ GeV, we show the $2\sigma$ limits achievable for
different combinations of angular and invariant mass cuts.
Experimental beam-pipe detection limitations motivated our ``weak''
angular cut of
$\theta _{C} = 0.95$; this cut itself removes a significant portion of
the peak region of phase space. A ``strong'' cut
$\theta_{C}$ was chosen so as to minimize
the contribution of the non-sensitive peak, whilst maximizing the total
cross-section. The
choices of $\theta_{C} = 0.7$ and $\theta_{C} = 0.9$ were taken as
representing the extremes
of optimizing the two opposing requirements, $\theta_{C} = 0.9$ maximizes
statistics and $\theta_{C} = 0.7$ excludes essentially all of the peak
contribution.

We see that making a cut on the variables can significantly improve
the limits for $\Delta \kappa < 0$, but has little effect for
$\Delta \kappa >0$. Also, the most restrictive combined cut of
$\theta_{C}=0.7$ and $100<M_{\mu \tau}<350$,
despite a sizable decrease in cross-section, gives the tightest constraint
on $\Delta \kappa < 0$ (but not for $\Delta \kappa > 0)$. It seems
that the improvement in physical sensitivity more than compensates for the
loss of statistics.

The other channels, $\mu^{+} e^{-},~\mu^{+} \mu^{-}$,
and $e^{+} e^{-}$,
because of the extra complexity due to the additional diagrams,
do not as cleanly provide the possibility of sensitivity enhancing cuts.
The additional $\kappa_{V}$
dependent diagrams can contribute
to the very regions of phase space that we previously considered
excluding; a cut to exclude these regions of phase space can therefore
be counter productive, with consequently no significant improvement in
achievable coupling limits.

The bounds quoted in Table I are for a
center of mass energy of 500 GeV. The sensitivity to $\kappa $ is generally
more evenly distributed over the available phase space at the higher
energy of 1 TeV.
Although the effect of the neutrino propagator is enhanced
at higher energies,
the contribution of this diagram becomes less significant as
$\sqrt{s} \gg 2M_{W}$. There is therefore less motivation for excluding
these angular regions to improve the limits on the $\kappa$
couplings.

\section{Detection Potential for $CP$-odd couplings}
\label{second}

We now extend our investigation to consider $CP$-odd couplings.
The $CP$ violating form factors $f_{4}, \tilde{\kappa}$, and $\tilde{\lambda}$
can, depending on the model
and the kinematics, have both real and imaginary parts. The effects
of the two different possibilities can be separated by
examining the consequences of the $CPT$ theorem and the unitarity
condition. The $CPT$ theorem postulates the invariance
\begin{equation}
<f|M|i>=<CPT(i)|M|CPT(f)>
\end{equation}
where $|CPT(j)>$ represents the state $|j>$ transformed by
$CPT$. It is very difficult
to directly check this symmetry because it requires the interchange
of initial and final states. It is more convenient to define a pseudo
time reversal transformation, $\tilde{T}$, that transforms the
kinematic observables of the initial and final state, as does $T$, but
does not interchange the initial and final states.

In the Born approximation, unitarity of the S-matrix implies that the
transition matrix, $M$, is hermitian. Thus, in the Born approximation
$M$ satisfies
\begin{displaymath}
M_{if}=M_{fi}^{*}
\end{displaymath}
And so, with a hermitian transition matrix, the $CPT$ theorem reduces to
\begin{equation}
<f|M|i>=<CP\tilde{T}(f)|M|CP\tilde{T}(i)>^{*}
\end{equation}
The $CPT$ theorem therefore provides a check on the hermiticity of the
transition matrix $M$. Non-hermiticity of $M$, which is due to contributions
beyond Born in which intermediate states can be on-shell, will manifest
itself in violations of $CP\tilde{T}$.

We approximate our full four lepton production processes by W pair
production for the purpose of discussion.
If we define $A_{\lambda,\overline{\lambda}}$ as the tree
level SM contribution to the transition matrix (with basis (-,0,+), the
helicities of the W's) and
$\delta A_{\lambda,\overline{\lambda}}$ as the
deviation due to the $CP$ violating couplings,
$\delta A_{\lambda,\overline{\lambda}}$ can be compactly expressed as
\cite{chang}
 \begin{equation}
\delta A_{\lambda,\overline{\lambda}} =
\left[\begin{array}{ccc}
-i(\beta^{-1}\tilde{\kappa} + 4\gamma^2 \beta \tilde{\lambda})&
-i\gamma(f_{4}+\beta^{-1}\tilde{\kappa}) & 0\\
-i\gamma(-f_{4}+\beta^{-1}\tilde{\kappa}) & 0 &
 i\gamma(f_{4}+\beta^{-1}\tilde{\kappa}) \\
0 & i\gamma(-f_{4}+\beta^{-1}\tilde{\kappa}) &
i(\beta^{-1}\tilde{\kappa} + 4\gamma^2 \beta \tilde{\lambda})
\end{array}     \right]
\end{equation}
The coefficients
in the above are $\gamma = \sqrt{s}/2m_{W}$ and
$\beta^{2} = 1- \gamma ^{-2}$.

Under the $CP\tilde{T}$ transformation, we must have
\begin{displaymath}
\delta A_{\lambda,\overline{\lambda}}
\stackrel{CP\tilde{T}}{\rightarrow}
\delta A_{-\overline{\lambda},-\lambda}^{*}
\end{displaymath}
which requires $f_{4}, \tilde{\kappa}$, and $\tilde{\lambda}$ to be real.
An imaginary component of the coupling parameters will
break the $CP\tilde{T}$ symmetry, and so parametrizes
the non-hermiticity of the transition matrix, the hallmark of
beyond Born final state interactions. Such effects are small in a
weakly coupled theory such as the SM, and so in what follows we
concentrate on the case where all form factors are real.

Of the three $CP$-odd couplings, $\tilde{\kappa}$ and $\tilde{\lambda}$
are $C$-even and $P$-odd, $f_{4}$ is $P$-even and $C$-odd. Thus a non-zero
$f_{4}^{\gamma}$ is forbidden by electromagnetic gauge invariance.
With $f_{4}^{\gamma}=0$, the existence of a non-zero $f_{4}^{Z}$ would
imply that the W boson interactions intrinsically violate the SU(2)
weak-isospin symmetry. There is, however, good empirical evidence for the
validity of this symmetry. Accordingly, we assume also that
$f_{4}^{\gamma}=f_{4}^{Z}=0$
and restrict ourselves to the couplings $\tilde{\kappa}_{\gamma},
\tilde{\kappa}_{Z},\tilde{\lambda}_{\gamma}$, and $\tilde{\lambda}_{Z}$
in the following Sections.
Also, in the analysis of these $CP$-odd couplings, we restrict the $CP$-even
couplings to their SM values, $\kappa_{V}=1$ and $\lambda_{V}=0$.

\subsection{Detection Limits on $\tilde{\kappa}_{V}$ and $\tilde{\lambda}_{V}$}
\label{detlim}

\subsubsection{Total Cross-Section Measurements}

We determine detection limits on $\tilde{\kappa}_{V}$ and
$\tilde{\lambda}_{V}$ at two center-of-mass energies, 500 GeV and 1 TeV.
For each of the four different types of four lepton channels, at each of the
two energies, we fit parabolas
to the dependence of the total cross-section on each of the couplings
$\tilde{\kappa}_{\gamma},\tilde{\kappa}_{Z},\tilde{\lambda}_{\gamma}$, and
$\tilde{\lambda}_{Z}$.
The Standard Model cross-section $\sigma_{SM}$, multiplied
by our assumed integrated
luminosity of 50 fb$^{-1}$, determines the expected number of events N, about
which we assume a normal distribution.
The detection limits, representing the magnitude of anomalous coupling
required to give a $2\sigma$ deviation in the number of events, are
listed below in Tables II and III.

The constraints on $\tilde{\kappa}_{V}$ from Table II
are quite loose compared to the predicted bounds on
the $CP$-even $\kappa_{V}$. The limits on $\kappa_{V}$
approach the few percent level at the higher
energy of $\sqrt{s}=1$ TeV.
The difference in sensitivity between the $CP$-even and $CP$-odd couplings
can be attributed in part to the relative phase between the $CP$-odd and
SM contributions.
This phase difference ensures that interference between the two is minimal;
thus the $CP$-odd couplings contribute predominantly to first order
quadratically, whereas the contributions of the non-standard $CP$-even
couplings
can interfere with the Standard Model amplitude.
The $\tilde{\lambda}_{V}$ terms also contribute
to first order dominantly quadratically; they however have a much
better high energy behaviour. They rise with energy like
$s$, and not $\sqrt{s}$, as do the $\tilde{\kappa}$ terms. The limits on
$\tilde{\lambda}_{V}$, as listed in Table III, are consequently
much tighter than those for $\tilde{\kappa}_{V}$, Table II.
Indeed, some of the limits on $\tilde{\lambda}_{V}$ from the different
channels are of the same scale as
predicted for these couplings by the various ``beyond-Standard'' models
\cite{hema,liu},
specifically, those from the $\mu^{+} \tau^{-}$ process at 1 TeV. Also,
they approach the level of precision predicted necessary by neutron electric
dipole moment measurements. Of course, these limits for $\tilde{\lambda}_{V}$
are dependent on the choice of the scaling parameter for the
$\tilde{\lambda}_{V}$ term in Eq. \ref{lag} (we chose $m_{W}^{2}$).
A choice of $\Lambda = 1$ TeV, reflecting the scale of possible ``new''
physics, sometimes suggested as being more appropriate, would
weaken the bounds on $\tilde{\lambda}_{V}$ by 2 orders of magnitude.

\subsubsection{Improved Limits from $\chi^2$ and Maximum Likelihood Analyses}
\label{chilim}

The sensitivity to the $CP$-odd couplings shows the same phase space
localization as was demonstrated for the $CP$-even couplings, with the
regions of least sensitivity contributing a large portion of the total
cross-section. We can account
for both the high and low sensitivity regions, and do so in a manner
which optimizes both statistics and sensitivity,
through a $\chi ^2$ \cite{choi} or  maximum likelihood analysis
\cite{bark,miki}.
We present briefly the basic principles of the two analyses and the
scale of the limits achievable with them.

To demonstrate the principle of the $\chi ^2$ analysis, we consider a
distribution where the sensitivity to $\tilde{\kappa}_{V}$ and
$\tilde{\lambda}_{V}$ is small where the differential cross-section is large,
and large where the differential cross-section is small (such as the angular
variable $\theta _{\tau^{-}}$).
We divide the phase space into bins, the choice of number and size of the bins
roughly determined by the regions of different sensitivity. We then
define our $\chi ^2$ test variable as a sum over these bins
\begin{displaymath}
\chi ^2=\sum_{i}^{n}\left[\frac{(X_i - Y_i)^2}{\Delta_{i}^{2}}\right]
\end{displaymath}
where
\begin{displaymath}
X_i = \left(\frac{d\sigma_{NSM}}{d\cos \theta_{l^-}}\right)_i,
Y_i = \left(\frac{d\sigma_{SM}}{d\cos \theta_{l^-}}\right)_i
\end{displaymath}
and $\sigma_{NSM}$ and $\sigma_{SM}$ are the anomalous and standard
cross-sections, respectively.
$\Delta_{i}^{2}$ combines both statistical and systematic errors for
the particular bin.
\begin{displaymath}
\Delta_{i}^{2}=(\Delta_{i}^{stat})^{2}+(\delta^{sys}Y_i)^2
\end{displaymath}

Where the cross-section is large, the sensitivity, and consequently the
``variance'' $(X_{i} - Y_{i})^{2}$, is small. However, if $Y_{i}$
is large, then the statistics should be improved, and $\Delta_{i}^{2}$
will also be small. Conversely, for regions where the sensitivity is large;
so also will $\Delta_{i}^{2}$ be, due to the poorer statistics. By summing
over these different regions, $\chi^{2}$ gives a more accurate estimate
of the deviation of the non-standard cross-section from the standard.

$\chi ^2$ was calculated from the
$\theta_{l^-}$ distribution as a sum over the following three bins,
$(-.95 \rightarrow -.5),(-.5 \rightarrow 0)$ and $(0 \rightarrow .95)$.
With 2 degrees of freedom, the 95\% C.L. corresponds to a $\chi^{2}$ of 6.0.
The systematic error was taken as $5\%$.

We determined that the limits from such a $\chi^2$ analysis are
generally improved by approximately 10$\%$, compared to those obtained
from an analysis of the total cross-section.

Even greater improvements are possible through a maximum likelihood analysis.
As an example, consider the $\tilde{\kappa}_{V}$ dependence of the
$\mu \tau$ process. We consider the two dimensional differential cross-section
\begin{displaymath}
\frac{d^2 \sigma}{d(\cos \theta _{\tau^{-}})d(M_{\mu \tau})}
\end{displaymath}
where $\theta _{\tau^{-}}$ and $M_{\mu \tau}$ are as defined previously. We
divide this distribution into bins and define the measured and
expected number of events in a given bin as $r_i$ and $\mu_i(\tilde{\kappa})$
respectively. If we assume that the number of events in each bin follows
a Poisson
distribution with a mean of $\mu_i$, then the probability of measuring
$r_i$ events is given by
\begin{displaymath}
y_{i}(\mu_{i}(\tilde{\kappa}),r_{i}) = e^{-\mu_{i}(\tilde{\kappa})}
	\frac{\mu_{i}(\tilde{\kappa})^{r_{i}}}{r_{i} !}
\end{displaymath}
and we define a likelihood function over all the bins as
\begin{displaymath}
L(\tilde{\kappa}) = \ln \left( \prod_{i}  e^{-\mu_{i}(\tilde{\kappa})}
	\frac{\mu_{i}(\tilde{\kappa})^{r_{i}}}{r_{i} !} \right)
\end{displaymath}
and our measure of deviation from the SM as
\begin{eqnarray*}
\Delta L & = & L(\tilde{\kappa}) - L(\tilde{\kappa}_{SM}) \\
\Delta L & = & \sum_{i} \left[ -(r_{i} - \mu_{i}(\tilde{\kappa}_{SM}))
	   + r_{i}(\ln r_{i} - \ln  \mu_{i}(\tilde{\kappa}_{SM}))\right]
\end{eqnarray*}
where  $L(\tilde{\kappa}_{SM})$ is the likelihood function
$L(\tilde{\kappa})$ with the couplings taking their SM values.

We divide the $\cos \theta_{\tau^{-}}$ and $M_{\mu \tau}$ distributions
into 5 bins each. We show in Fig. 3
the $\Delta L = 2$ contour (corresponding to a $2\sigma$ significance
level) in $\tilde{\kappa}_{\gamma}$ and $\tilde{\kappa}_{Z}$ for the
$\mu^{+}\tau^{-}$ process at $\sqrt{s}$ of 500 GeV.
We extract limits from this contour of
\begin{eqnarray*}
-0.10 < & \tilde{\kappa}_{\gamma} & < 0.10 \\
-0.11 < & \tilde{\kappa}_{Z}      & < 0.12
\end{eqnarray*}
which are a significant improvement on the constraints achievable
from a simple total cross-section measurement. Similarly, the limits for the
other parameters and other processes were generally improved by approximately
$50 \%$ relative to the total cross-section limits through such an analysis.

\subsection{$CP$ Asymmetries as Indicators of $CP$ Violation}
\label{asym}

A more intuitive means of identifying a $CP$-violating
contribution is, rather than looking for non-standard effects in
cross-sections,
to search directly for evidence of the breaking of the
symmetry. Even a small amount of $CP$ violation in the triple boson vertex
could in principle produce clear experimental signatures. These
signatures could consist of asymmetries, for instance, in the
numbers of events between two $CP$-conjugate states. As an example, in the
process $e^{+}e^{-} \rightarrow W^{+}W^{-}$, a difference between the numbers
of $W^{+}$ bosons emitted in the ``forward'' direction, and the number of
$W^{-}$ bosons in the ``backward'' direction, as it distinguishes
between these $CP$-conjugate states, would indicate the breaking of $CP$.
Different types of  observable asymmetries have been suggested as possible
indicators of $CP$-violation \cite{rindani}; these include width asymmetries,
partial rate asymmetries such as energy and angular asymmetries,
and $CP$-odd correlations.

We search for non-zero asymmetries in certain measurable
observables as evidence for a $CP$-violating contribution to the TBV.
The observables we consider are defined in terms
of the final state charged lepton and anti-lepton momenta and/or polar and
azimuthal angles, and thus avoid any ambiguity from neutrino non-detection.

We define the polar and azimuthal angles of the final state lepton
and anti-lepton $(\theta,\phi)$ and $(\overline{\theta},\overline{\phi})$
through the momentum parametrization
\begin{eqnarray*}
\vec{p}=|\vec{p}|(\sin \phi \sin \theta,\sin \phi \cos\theta,\cos\phi)
\nonumber \\
\vec{q}=|\vec{q}|(\sin \overline{\phi} \sin \overline{\theta},
\sin \overline{\phi} \cos \overline{\theta},
\cos \overline{\phi})
\end{eqnarray*}
where $\vec{p}$ and $\vec{q}$ are the three-momenta of the outgoing charged
lepton and charged antilepton, respectively. The $CP$ operation
results in the following transformation amongst the angular variables:
\begin{equation}
(\theta,\phi,\overline{\theta},\overline{\phi})
\stackrel{CP}{\leftrightarrow}
(\pi+\overline{\theta},\pi-\overline{\phi},\pi+\theta,\pi-\phi)
\end{equation}

If the transition matrix $M$ is hermitian, then $CP\tilde{T}$
invariance gives the following relation:
\begin{displaymath}
M_{\sigma\overline{\sigma};\lambda\overline{\lambda}}
= M^{*}_{-\overline{\sigma},-\sigma;-\overline{\lambda},
-\lambda}
\end{displaymath}
The combined $CP\tilde{T}$
is therefore equivalent to the following transformation amongst
the angular variables
\begin{equation}
(\theta,\phi,\overline{\theta},\overline{\phi})
\stackrel{CP\tilde{T}}{\leftrightarrow}
(\pi-\overline{\theta},\pi-\overline{\phi},\pi-\theta,\pi-\phi)
\end{equation}

We can now classify angular distributions according to their
behaviour under $CP$ and $CP\tilde{T}$. In light of the previous discussion,
we concentrate on $CP$-odd and $CP\tilde{T}$-even variables and restrict
ourselves to the real parts of $\tilde{\kappa}_{V}$ and
$\tilde{\lambda}_{V}$.

We examined asymmetries in the following variable, defined here as $S$
\cite{hag}

\begin{eqnarray*}
S & = & \sin\phi \sin\theta +\sin \overline{\phi} \sin \overline{\theta}
\end{eqnarray*}

If $CP$ is a valid symmetry, then a differential cross-section in $S$
would necessarily be symmetric about $S$=0.
The presence of $CP$ violating couplings
such as $\tilde{\kappa}_{V},\tilde{\lambda}_{V}$
will manifest itself in the loss of this symmetry.
We define an asymmetry  as
\begin{equation}
A^{S}_{CP}=\frac{\int d\sigma(S>0)-\int d\sigma(S<0)}
		{\int d\sigma(S>0)+\int d\sigma(S<0)}
\end{equation}

We can also determine the asymmetries in the individual helicity amplitudes,
defined analogously to that in the total cross-section.
Non-zero values for these helicity asymmetries combine to
produce a non-zero value for the total cross-section; it is therefore
possible for large oppositely signed asymmetries in the helicities
to combine to give a smaller net asymmetry in the total cross-section.

For $\tilde{\kappa}_{V}$ and $\tilde{\lambda}_{V}$ equal to 1, we show in
Table IV the asymmetries $A^{S}_{CP}$
in the
total cross-section for the lepton
channels $\mu^{+} \tau^{-},\mu^{+}\mu^{-}$, and $e^{+}e^{-}$. (Although
$\mu^{+} \tau^{-}$ is not a $CP$ eigenstate, having neglected lepton masses,
this channel effectively approximates the $CP$ invariant
$W^{+}W^{-}$ production. The
$\mu^{+}e^{-}$ channel, without its charge conjugate process $\mu^{-}e^{+}$,
is not $CP$ invariant; it also introduces TBV dependent diagrams additional to
those of $W^{+}W^{-}$). The $CP$-odd couplings contribute predominantly
to first-order quadratically; we therefore
restrict our study to positive anomalous couplings
$\tilde{\kappa}_{V},~~\tilde{\lambda}_{V} = 1.0$.

We notice that the asymmetries are of the same magnitude for each of the
different final state configurations, and that the higher energy does not
guarantee larger asymmetries. The typical magnitude of these values
of $\sim 10^{-3}$ agrees with the prediction of Mani {\it et al.} \cite{mani}.
These authors, instead of simple asymmetries, looked at expectation values of
$CP$-odd variables, and find A$\sim 10^{-3}$ for
$\tilde{\kappa} = \tilde{\lambda} = 0.1$. We also considered asymmetries in the
$CP$-odd, $CP\tilde{T}$-even variable
$\vec{k}_{2}\cdot(\vec{p}_{1}\times\vec{q}_{1})$ \cite{chang} where
$\vec{k}_{2}$
is the vector momentum of the incoming electron, $\vec{p}_{1}$ that of the
outgoing lepton,  and $\vec{q}_{1}$ that
of the outgoing antilepton. The results were similar to those for the variable
$S$.

For an asymmetry in the total cross-section to be measurable, we require
that the number of asymmetrical
events $\Delta N$ exceed the fluctuations about the total number of events.
Thus our significance requirement for these asymmetries is:
\begin{eqnarray*}
\Delta N   & > & \delta N \\
A \sigma L & > & \sqrt{\sigma L} \\
A          & > & (\sigma L) ^{-\frac{1}{2}}
\end{eqnarray*}
If we take a typical value for the cross-section of $\sigma \sim 0.1$ pb ,
and with our assumed integrated luminosity of 50 fb$^{-1}$, this
significance level is approximately 1.5 \%. From this quick calculation, it
seems that an
asymmetry in the total cross-section will most likely be below the
statistical significance level, and therefore unresolvable. The situation
is even less encouraging if, instead of
$\tilde{\kappa}_{V}=\tilde{\lambda}_{V}=1.0$, we explore the expected
asymmetries for more realistic magnitudes for the anomalous couplings,
ie. $\tilde{\kappa}_{V}$ and $\tilde{\lambda}_{V}$ at their detection limits.

Although it appears that any asymmetry in total cross-section will be
below the level of statistical significance, asymmetries in the helicity
amplitudes that contribute to the total cross-section might be detectable.
An asymmetry in the total
cross-section is a result of a combination of asymmetries in the contributing
helicity amplitudes, each weighted by their appropriate helicity
cross-sections.
Since the different helicity amplitudes have contributions from
different TBV diagrams, with consequently different sensitivity to anomalous
couplings, the asymmetries in these helicity amplitudes can differ in
magnitude and sign. In principle therefore, large but opposing asymmetries
can cancel each other to produce a smaller resultant asymmetry in the total
cross-section. We demonstrate this idea by considering the
``differential asymmetries''. We define this differential asymmetry
$\chi(S)$ as
\begin{equation}
\label{chi}
\chi(S) = \frac{\left(\frac{d\sigma}{d(S)}
	      - \frac{d\sigma}{d(-S)}\right)}{\sigma}
\end{equation}
and consider both total and helicity distributions.
We show in Fig. 4 the distributions
in $\chi$(S) for the $\mu^{+} \tau^{-}$ channel at $\sqrt{s}=500$ GeV for
non-standard $\tilde{\kappa}_{Z}=1.0$. As was previously mentioned, we use the
$\mu^{+} \tau^{-}$ channel to approximate the $W^{+}W^{-}$ process.
We show also the approximate
statistical error bars about the SM expectation of $\chi$(S)=0, calculated
from the differential cross-section $d\sigma/dS$. For SM couplings,
$\chi(S)$ vanishes for each helicity individually.

We see that while the asymmetry in the total cross-section, as it is bounded
by the error bars, would be below the statistical significance level and so
unresolvable; that in the $(+-+-)$ helicity
amplitude might however be measurable. A polarized beam facility, by
separately generating the different helicity components, might be able to
measure these helicity asymmetries and so access this phenomena.

\section{Conclusions}
\label{con}

We have presented an analysis of the sensitivity to the WWV coupling parameters
through measurement of the processes $e^+ e^- \rightarrow l^+ \nu l^{-\prime}
\overline{\nu}$ at center-of-mass energies of 500 GeV and 1 TeV.
The limits on the $CP$-even coupling $\kappa_{V}$ are of the order of $1\%$ at
the higher energy, with slightly looser bounds achievable at the lower energy.
These bounds might be significantly improved at a polarized beam
facility. Because the individual helicity amplitudes can have very different
dependence on the TBV, in form and magnitude, the limits obtainable from an
analysis of an individual helicity amplitude can be complementary to
those from the other helicities and from a total
cross-section analysis. Consequently, combining the limits from the
different helicity amplitudes can, for certain of the four lepton final state
configurations, significantly tighten the constraints on the
anomalous couplings. Smaller improvements are also possible through
suitably cutting on the phase space of certain experimental variables.
The sensitivity to anomalous couplings is not always evenly distributed
over these variable's distributions; a cut to isolate the regions
of high sensitivity and exclude those of low sensitivity can improve the
achievable TBV bounds. These improvements can be of the order of a factor
of two for the $\mu^{+} \tau^{-}$ process at the lower energy.

The limits on the $CP$-odd coupling $\tilde{\lambda}_{V}$, especially at the
higher energy, approach the level of precision predicted necessary by
neutron electric dipole moment measurements; the limits on the $CP$-odd
variable $\tilde{\kappa}_{V}$ are much looser.
These limits can be improved by accounting for the uneven localization of
TBV sensitivity in certain experimental variables through a $\chi ^{2}$
analysis or a maximum likelihood fit.

An explicitly $CP$-violating vertex contribution is unlikely to produce
measurable asymmetries in the total cross-section.
Asymmetries in certain $CP$-odd variables might however be measurable
in the component helicity amplitudes; a polarized beam facility would
be required for this measurement.

\section*{Acknowledgements}

This work was funded in part by the Natural Sciences and Engineering Council
of Canada. The authors thank Mikulas Gintner and W.Y. Keung for
helpful discussions.

\newpage

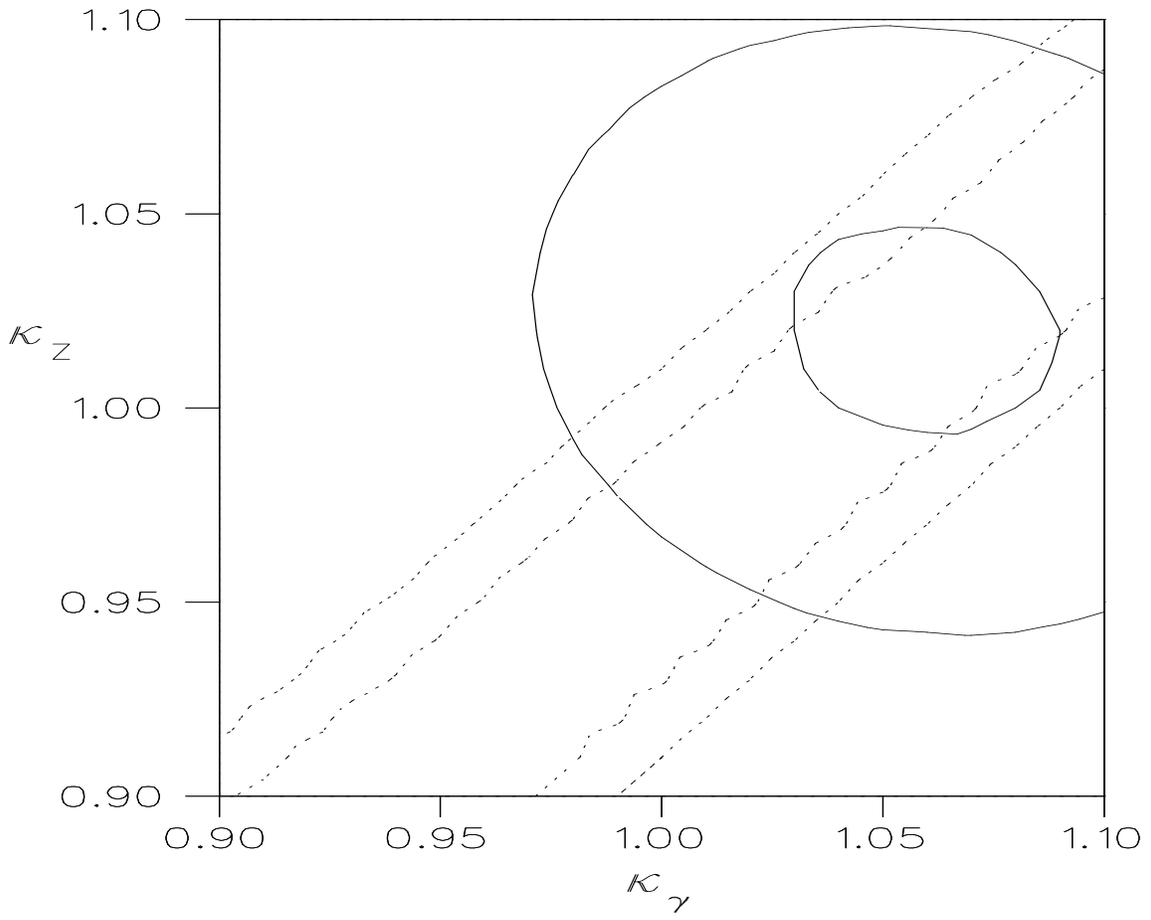
\begin{figure}
\caption[]
{$\pm 2\sigma$ contour plots for $(+-+-)$ helicity amplitude
(solid line) and $(-++-)$ helicity amplitude (dashed line) of
$\mu^{+} \tau^{-}$ channel at $\sqrt{s} = 500$ GeV.}
\end{figure}

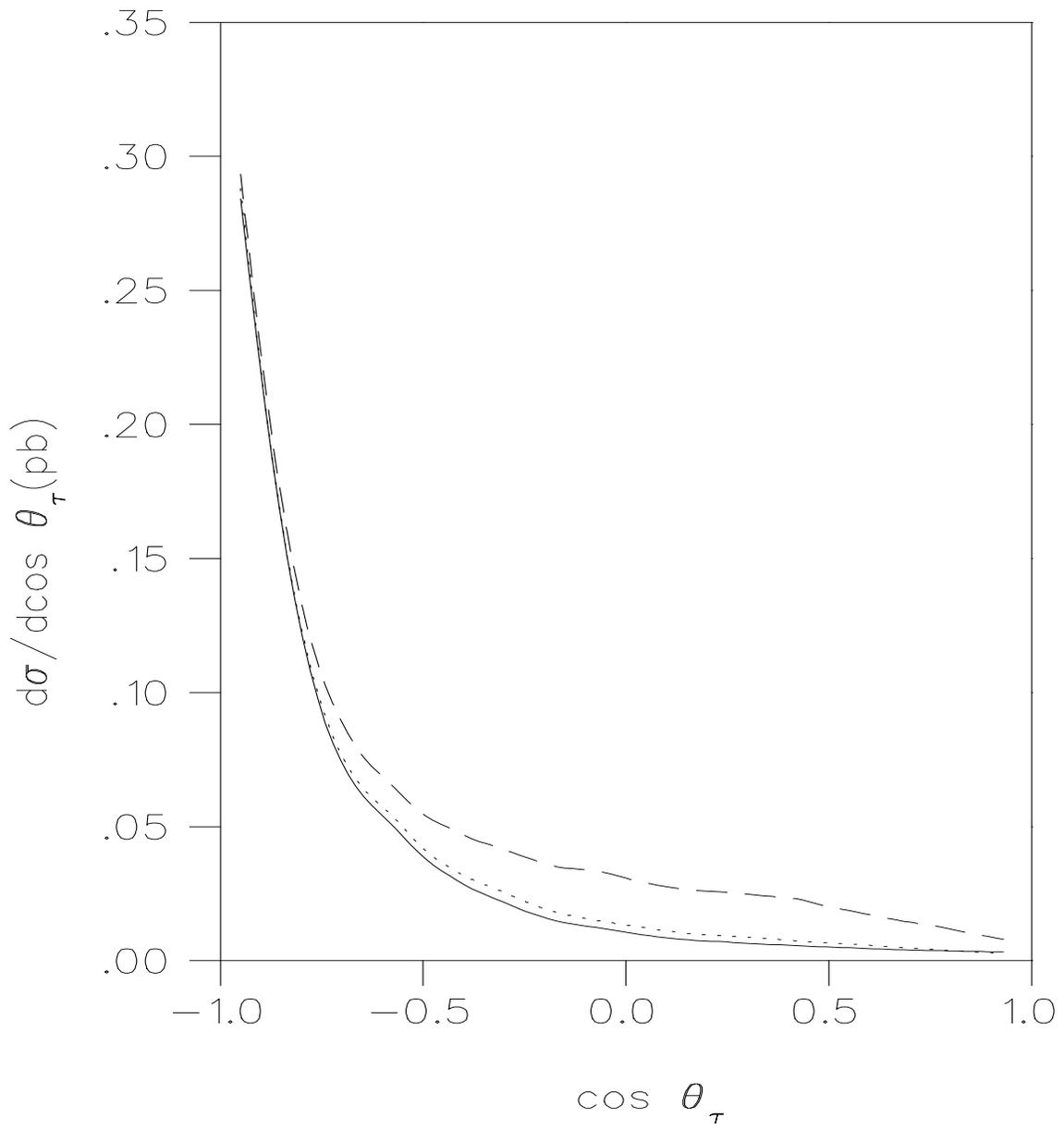
\begin{figure}
\caption[]{Differential cross-sections with respect to
(a) $\cos \theta_{\tau^{-}}$
for Standard Model(solid line),
$\Delta \kappa _{\gamma}=\Delta \kappa _{Z}=-0.1$(dashed line),
and $\Delta \kappa _{\gamma}=\Delta \kappa _{Z}=0.1$(dotted line).}
\end{figure}

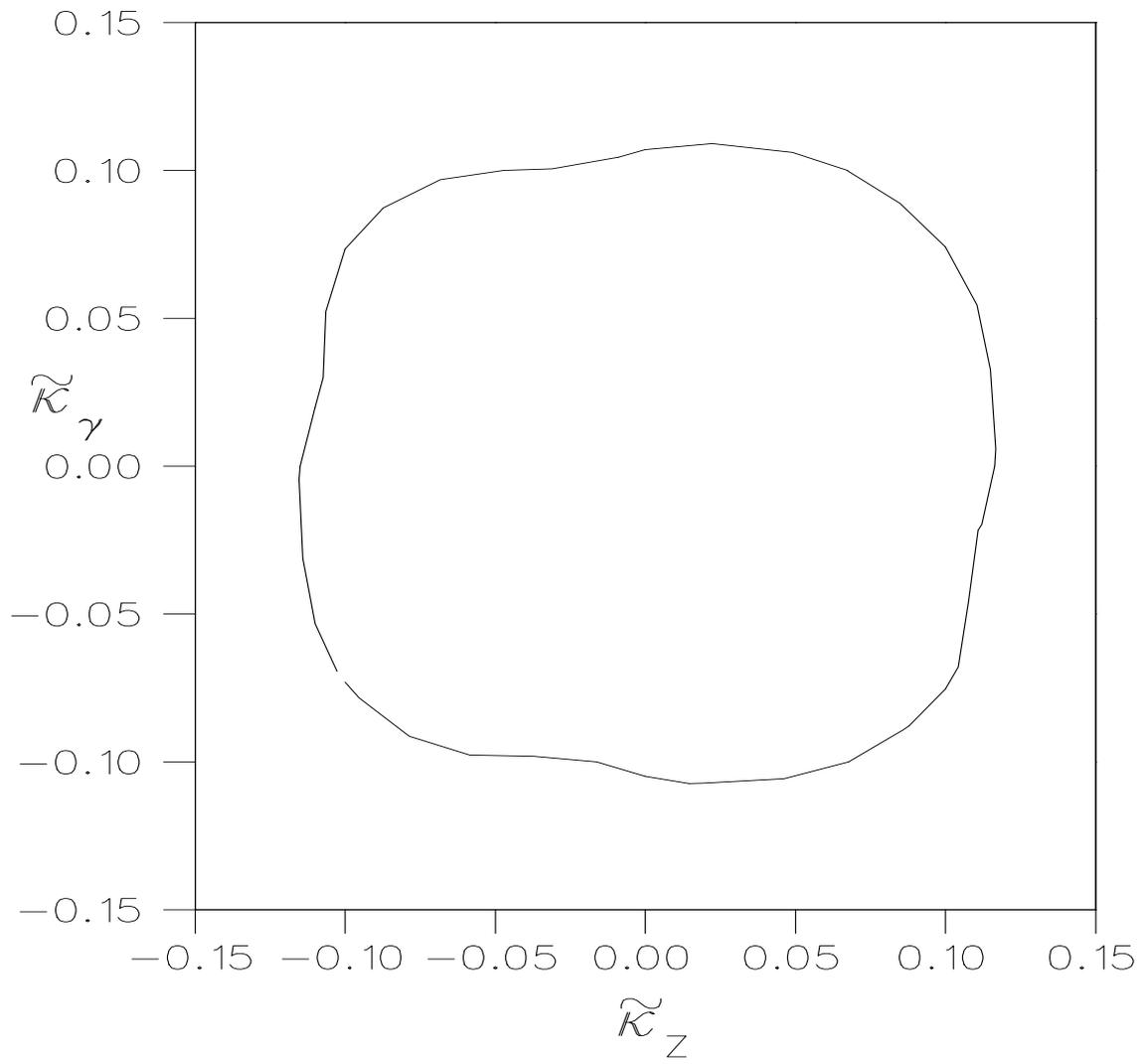
\begin{figure}
\caption[]{$\Delta L = 2$ contour in $\tilde{\kappa}_{\gamma}$ and
$\tilde{\kappa}_{Z}$ for maximum likelihood analysis for
$\mu^{+}\tau^{-}$ process at $\sqrt{s} = 500$ GeV.}
\end{figure}

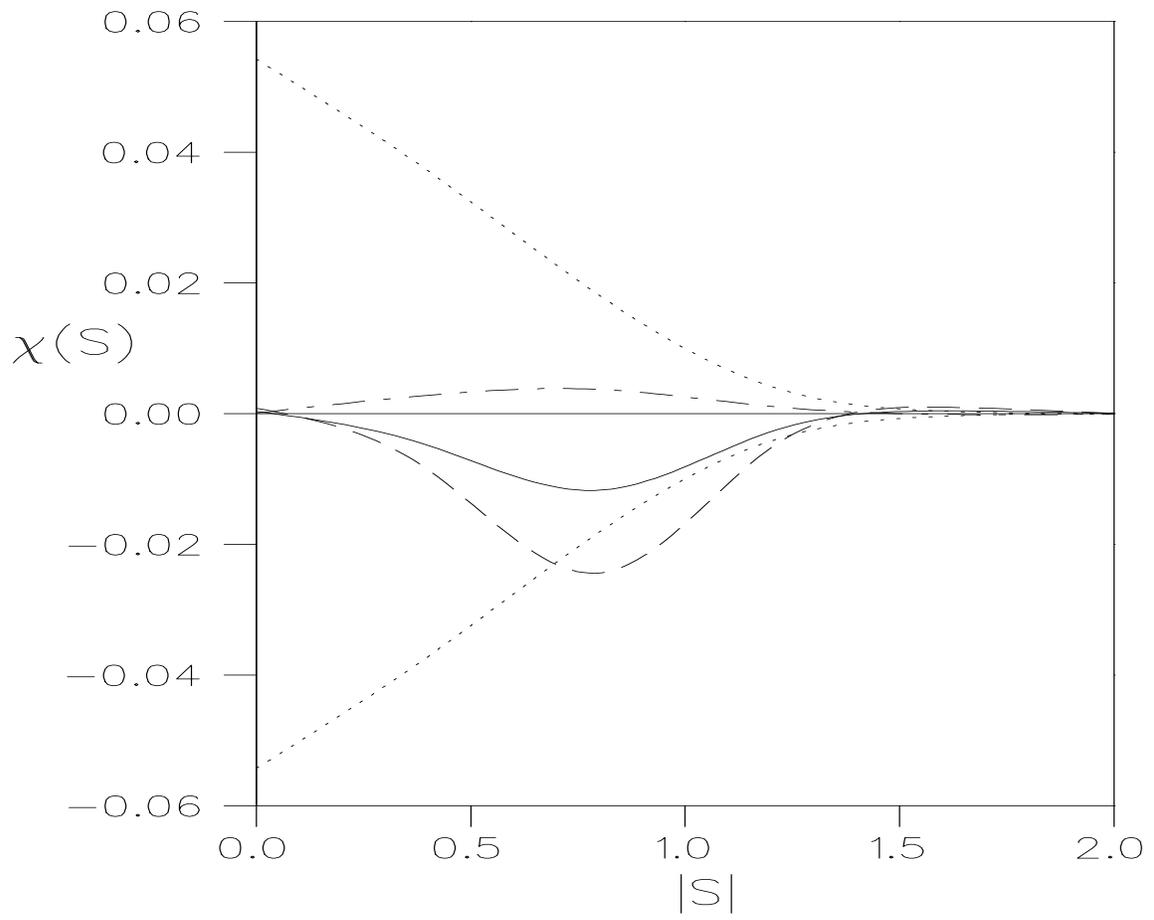
\begin{figure}
\caption[]{Distributions in $\chi$(S) for the $\mu^+ \tau^-$ process at
center-of-mass energy of 500 GeV with $\tilde{\kappa}_{Z}=1.0$.
The solid line corresponds to the
total cross-section, the dot-dashed to the $(-++-)$ amplitude, and the
dashed to the $(+-+-)$ amplitude. The asymmetries are divided by the
appropriate cross-section, either total or component. The dotted line
represents the statistical error bounds.}
\end{figure}

\newpage
\begin{center}
\begin{table}[htb]
\caption[$2\sigma$ limits for various cut combinations for
$\mu^+ \tau^-$ process at 500 GeV.]
{$2\sigma$ limits for various cut combinations for
$\mu^+ \tau^-$ process at $\sqrt{s} = 500$ GeV.}
\[
\begin{array}{|c|c|c|} \hline
\mbox{Cuts}      & \Delta \kappa < 0    & \Delta \kappa > 0\\ \hline
\theta_{C}= 0.95        & 1.5 \%             & 9.3 \%          \\
\theta_{C}=0.9         & 1.2 \%             & 9.3 \%          \\
\theta_{C}=0.7         & 0.8 \%             & 9.0 \%          \\
\theta_{C}=0.7,30<M_{\mu \tau}<430 & 1.0 \% & 8.5 \%          \\
\theta_{C}=0.7,100<M_{\mu \tau}<350& 0.79 \% & 9.0 \%          \\ \hline
\end{array} \]
\end{table}

\begin{table}[htb]
\caption{$2\sigma$ bounds on non-standard couplings $\tilde{\kappa}_{V}$}
\[
\begin{array}{|c|c|c|} \hline
\mbox{Process}& \sqrt{s}~(\mbox{GeV}) & \mbox{Sensitivity Limits}  \\ \hline
\mu \tau & 500  & -0.19<\tilde{\kappa} _{\gamma}<~0.18
             ~~     -0.16<\tilde{\kappa} _{Z}<0.16 \\ \cline{2-3}
\rule{0.5in}{0in}& 1000 & -0.13<\tilde{\kappa} _{\gamma}< 0.13
             ~~     -0.12<\tilde{\kappa} _{Z}< 0.12 \\ \hline
\mu e    & 500 & -0.17<\tilde{\kappa} _{\gamma}< 0.18
             ~~     -0.17<\tilde{\kappa} _{Z}< 0.18 \\ \cline{2-3}
\rule{0.5in}{0in}& 1000 & -0.14<\tilde{\kappa} _{\gamma}< 0.15
             ~~     -0.15<\tilde{\kappa} _{Z}< 0.15 \\ \hline
\mu \mu  & 500 & -0.24<\tilde{\kappa} _{\gamma}< 0.24
             ~~     -0.18<\tilde{\kappa} _{Z}< 0.17 \\ \cline{2-3}
\rule{0.5in}{0in}& 1000 & -0.21<\tilde{\kappa} _{\gamma}< 0.21
             ~~     -0.10<\tilde{\kappa} _{Z}< 0.10 \\ \hline
ee      & 500 & -0.17<\tilde{\kappa} _{\gamma}< 0.17
             ~~     -0.19<\tilde{\kappa} _{Z}< 0.19 \\ \cline{2-3}
\rule{0.5in}{0in}& 1000 & -0.15<\tilde{\kappa} _{\gamma}< 0.15
             ~~     -0.11<\tilde{\kappa} _{Z}< 0.11      \\ \hline
\end{array} \]
\end{table}

\begin{table}[htb]
\caption{$2\sigma$ bounds on non-standard couplings $\tilde{\lambda}_{V}$}
\[
\begin{array}{|c|c|c|} \hline
\mbox{Process}  & \sqrt{s}~(\mbox{GeV}) & \mbox{Sensitivity Limits}  \\ \hline

\mu \tau & 500 & -0.013<\tilde{\lambda} _{\gamma}< 0.012
              ~~-0.011<\tilde{\lambda} _{Z}< 0.011 \\ \cline{2-3}
\rule{0.5in}{0in}& 1000 & -0.00098<\tilde{\lambda} _{\gamma}< 0.00095
              ~~    -0.00090<\tilde{\lambda} _{Z}< 0.00087 \\ \hline
\mu e    & 500 & -0.0073<\tilde{\lambda} _{\gamma}< 0.00074
              ~~    -0.0070<\tilde{\lambda} _{Z}< 0.0068 \\ \cline{2-3}
\rule{0.5in}{0in}& 1000 & -0.00077<\tilde{\lambda} _{\gamma}< 0.00079
              ~~    -0.00084<\tilde{\lambda} _{Z}< 0.00081 \\ \hline
\mu \mu  & 500 & -0.015<\tilde{\lambda} _{\gamma}< 0.015
              ~~  -0.014<\tilde{\lambda} _{Z}< 0.014 \\ \cline{2-3}
\rule{0.5in}{0in}& 1000 & -0.0016<\tilde{\lambda} _{\gamma}< 0.0016
              ~~ -0.0014<\tilde{\lambda} _{Z}< 0.0014 \\ \hline
ee       & 500 & -0.011<\tilde{\lambda} _{\gamma}< 0.011
              ~~ -0.010<\tilde{\lambda} _{Z}< 0.0095 \\ \cline{2-3}
\rule{0.5in}{0in}& 1000 & -0.0013<\tilde{\lambda} _{\gamma}< 0.0013
              ~~  -0.0013<\tilde{\lambda} _{Z}< 0.0012  \\ \hline
\end{array} \]
\end{table}

\begin{table}[htb]
\caption{Asymmetries $A^{S}_{CP}(\times 10^3)$ in Total Cross-Section}
\[
\begin{array}{|c|c|c|c|c|c|} \hline
\mbox{Process} & \sqrt{s}~(\mbox{GeV})& \tilde{\kappa}_{\gamma}=1
                                     & \tilde{\kappa}_{Z}=1
                                     & \tilde{\lambda}_{\gamma}=1
                                     & \tilde{\lambda}_{Z}=1 \\ \hline

\mu \tau        & 500 & -2.1 & -3.3  & -1.9  & -1.8  \\ \cline{2-6}
\rule{.5in}{0in}        & 1000& 0.58  & 1.1  & -1.9  & -2.0  \\ \hline

\mu \mu         & 500 & -1.3  & -0.67 & 3.2  & 3.9  \\ \cline{2-6}
\rule{.5in}{0in}        & 1000& -0.14  & -1.7  & -1.1  & -1.3  \\ \hline

ee              & 500 & -2.2  & -2.5 & -3.7  & -4.1 \\ \cline{2-6}
\rule{.5in}{0in}        & 1000& 0.16  & 0.85  & -6.3  & -6.7  \\ \hline
\end{array} \]
\end{table}

\end{center}

\end{document}